\documentclass{mn2e}
\usepackage[utf8x]{inputenc}
\usepackage{float}
\usepackage{graphicx}
\usepackage{amssymb}
\usepackage{amsmath}
\usepackage[usenames,dvipsnames]{xcolor}
\usepackage{astrojournals}
\usepackage{multicol}
\usepackage{multirow}
\usepackage[]{hyperref}
\usepackage{enumerate}
\usepackage{natbib}
\usepackage{placeins}

\newcommand{\bicep}{\textit{BICEP}2}
\newcommand{\planck}{\textit{Planck}}
\newcommand{\litep}{\textit{Low-$\Omega_{\rm M}$}}
\newcommand{\wdmp}{\textit{WDM}$_\mathrm{2.6keV}$}

\newcommand{\mvir}{M_\mathrm{v}}
\newcommand{\rvir}{R_\mathrm{v}}
\newcommand{\rmax}{R_\mathrm{max}}
\newcommand{\vvir}{V_\mathrm{v}}
\newcommand{\vmax}{V_\mathrm{max}}

\newcommand{\msun}{M_\odot}
\newcommand{\lsun}{L_\odot}

\newcommand{\mpc}{\mathrm{Mpc}}
\newcommand{\kpc}{\mathrm{kpc}}
\newcommand{\hmsun}{h^{-1}\, M_\odot}
\newcommand{\hmpc}{h^{-1}\,\mathrm{Mpc}}
\newcommand{\hkpc}{h^{-1}\,\mathrm{kpc}}
\newcommand{\kms}{{\rm km} \, {\rm s}^{-1}}
\newcommand{\lcdm}{$\Lambda$CDM}
\newcommand{\vhalf}{V_{1/2}}
\newcommand{\rhalf}{r_{1/2}}

\newcommand{\om}{\Omega_{\rm m}}
\newcommand{\ol}{\Omega_\Lambda}
\title{Running with \bicep: Implications for Small-Scale Problems in CDM}
\author[S.\ Garrison-Kimmel et al.]{Shea Garrison-Kimmel$^1$\thanks{$\!$sgarriso@uci.edu},
  Shunsaku Horiuchi$^{1,2}$,
  Kevork N.\ Abazajian$^1$, \and
  James S.\ Bullock$^1$, 
  Manoj Kaplinghat$^1$ \\
  \noindent$\!\!$ $^1$Center for Cosmology, Department of Physics and Astronomy,
  University of California, Irvine, CA 92697, USA \\
  \noindent$\!\!$ $^2$McCue Fellow}
\begin{document}

 \pagerange{\pageref{firstpage}--\pageref{lastpage}} 
 \pubyear{2014}

\maketitle

\label{firstpage}
\begin{abstract} 
The \bicep\ results, when interpreted as a gravitational wave signal and 
combined with other CMB data, suggest a roll-off in power towards small 
scales in the primordial matter power spectrum. Among the simplest 
possibilities is a running of the spectral index.  Here we show that 
the preferred level of running alleviates small-scale issues within the 
\lcdm\ model, more so even than viable WDM models.  We use cosmological 
zoom-in simulations of a Milky Way-size halo along with full-box 
simulations to compare predictions among four separate cosmologies: a 
\bicep-inspired running index model ($\alpha_s = -0.024$), two 
fixed-tilt \lcdm\ models motivated by \planck, and a $2.6$~keV thermal WDM 
model.  We find that the running \bicep\ model reduces the central densities 
of large dwarf-size halos ($\vmax \sim 30 - 80 \, \kms$) and alleviates the 
too-big-to-fail problem significantly compared to our adopted 
\planck\ and WDM cases.  Further, the \bicep\ model suppresses the count of small
subhalos by $\sim50\%$ relative  to \planck\ models, and yields a significantly 
lower ``boost" factor for dark matter annihilation signals.  Our findings highlight 
the need to understand the shape of the primordial power spectrum in order to correctly 
interpret small-scale data.
 \end{abstract}

\begin{keywords}
dark matter -- cosmology: theory -- galaxies: halos -- Local Group
\end{keywords}

\section{Introduction}
\label{sec:intro}

The discovery of the cosmic microwave background (CMB) and measurements 
of its temperature anisotropy have lead to a standard cosmological
model consisting of a flat universe dominated by cold dark matter and a 
cosmological constant that drives accelerated expansion at late times \citep[e.g.,][]{PlanckCosmo}. 
Inflation extends this standard cosmology by positing an earlier period 
of rapid exponential expansion that sets the initial conditions for the 
hot big bang; this period alleviates a number of ``fine-tuning" problems,
but lacked supporting observational evidence. Recently, however, the 
\bicep\ experiment reported the detection of primordial B-modes in the 
CMB \citep{BICEP22,BICEP21}. One explanation for this 
signal is the stochastic background of gravitational waves generated by 
inflation, providing potentially the first direct evidence for an 
inflationary phase in the early Universe. This explanation will have 
to be verified by other experiments and in other frequencies. For the rest of this
paper, we will assume this explanation is correct as we await confirmation by 
other experiments and in other frequency bands.~\footnote{In this regard, note 
that there has been concern that foreground contamination could have 
affected this measurement \citep[e.g.][]{Liu2014}.}

The tensor-to-scalar ratio measured by \bicep, $r=0.20^{+0.07}_{-0.05}$ 
($68$\% confidence-interval), is at face value inconsistent with the limit 
quoted from a combination of \planck\ \citep{PlanckCosmo,Ade:2013uln}, 
{\it SPT} \citep{Hou:2012xq}, {\it ACT} \citep{Das:2013zf}, and {\it WMAP} 
polarization \citep{Hinshaw:2012aka} data: $r<0.11$ at $95$\% 
confidence.\footnote{
	As noted by \citet{Audren2014}, however, the measured tension may
	be significantly reduced ($\sim1.3\sigma$) by assuming identical 
	values for the pivot scale and the tensor spectral index in both 
	analyses, effectively raising the upper limits on the running 
	measured by \planck.}
 However, these pre-\bicep\ limits assumed a constant spectral index $n_s$ for
scalar fluctuations in the primordial power spectrum.
The discrepancy could be explained by a nontrivial primordial power 
spectrum, one that deviates from a pure power law 
\citep[e.g.,][]{Hazra:2014aea}; suppressing the large-scale scalar power 
spectrum relative to that expected in a constant spectral index model 
allows for a larger contribution from tensor modes to the
temperature-temperature anisotropy $C_l^{TT}$ at large scales .
\cite{KnottedSky2} explored several scenarios including a running
spectral index, a cutoff in the spectrum, and a break in the power
spectrum, finding evidence for a negative running index 
\citep[see also][]{McDonald:2014kia,Ashoorioon2014} or for a broken spectrum.  Of these
possibilities, the running spectral index is arguably the simplest,
and we focus on the small-scale implications of this solution for the
remainder of this work.  More generally, however, our results explore the
possible implications of non-trivial primordial power spectra on galaxy 
formation.  Here we specifically show that viable deviations from power-law 
primordial power spectrum can have a significant impact on 
important questions facing $\Lambda$CDM today.

Any modifications to the primordial power spectrum and cosmological parameters 
will manifest itself in the formation and evolution of large-scale 
structure.  On large scales, the standard $\Lambda$CDM cosmology provides 
an excellent model for the observed Universe \citep[][]{Ho:2012vy,Hinshaw:2012aka}; 
any changes that compromise this success would thus be a sign of an 
inconsistent scenario.

On the other hand, discrepancies currently exist between the $\Lambda$CDM 
paradigm and the observed Universe on smaller scales. Examples include the 
``core/cusp problem,'' where dissipationless $N$-body simulations in 
$\Lambda$CDM predict a rising dark matter density with smaller radius 
$\rho \propto r^{-1}$, in contrast to observations that show a core-like 
profile at small radii \citep{Flores:1994gz,Moore1999}.  The discrepancy is 
seen in low-surface brightness (LSB) galaxies 
\citep{Simon2005,Donato2009,deNaray:2010zw,Oh:2010ea}, 
but also seems to appear in lower luminosity dwarf spheroidal 
(dSph) galaxies
	\footnote{We note that the density profiles of dSphs are currently a matter of some debate \citep[e.g.][]{Breddels2013}.} 
\citep{Walker:2011zu,Agnello2012,Amorisco2012}.  A second discrepancy is that 
the count of known satellite galaxies around the Milky Way is much smaller than 
the count of subhalos expected to be massive enough to form stars 
\citep[][the ``missing satellites problem"]{Klypin1999,Moore1999}.  Independently, 
it has also been shown that the central densities of dSphs are significantly 
lower than predicted by dissipationless $\Lambda$CDM simulations, dubbed 
the ``too-big-to-fail problem'' \citep[TBTF; ][]{MBK2011,MBK2012}.   The severity
of TBTF remains an active debate in the literature, with some authors pointing
out that a reduced MW mass would effectively eliminate the problematic halos 
\citep[e.g.][]{Wang2012,Cautun2014} and others arguing that baryonic processes, 
such as reionization, supernovae feedback, tidal interactions, and ram pressure 
stripping, may reduce the central densities of simulated dwarf halos \citep[e.g.][]{Bullock2000,Somerville2002,Pontzen2012,Zolotov2012,
BrooksZolotov2012,Arraki2012,Gritschneder2013,Garrison2013,Amorisco2013feedback,
DelPopolo2014,Sawala2014,Pontzen2014}.

Quantitatively,  the magnitude of these small-scale problems and the degree to which
feedback and other baryonic processes can operate to solve them depend on the underlying 
power spectrum and cosmological parameters, which fundamentally affect the collapse 
times and central densities of dark matter halos.  For example,  
\citet{Zentner2002, Zentner2003} showed that non-trivial primordial power spectra of 
the type expected in basic inflation models can alleviate many of the small-scale 
problems faced by $\Lambda$CDM, and used semi-analytic models to show that running 
at the level of $\alpha_s \simeq -0.03$ can reduce discrepancies significantly.  Later, 
using numerical simulations,  \citet{Polisensky:2013ppa} showed that differences in 
best-fit $\sigma_8$ and $n_s$ values between WMAP data releases impact small-scale 
predictions in important ways.  The implication is that changes that follow from the 
\bicep\ results can affect the magnitude of small-scale discrepancies significantly.  
Similarly, imposing a free-streaming cutoff in the initial power spectrum (e.g. from 
warm dark matter, WDM, or from a non-trivial inflation model) may also aid in resolving 
problems \citep{Kamionkowski2000,Zentner2003,Kaplinghat2005,Lovell2013,Schneider2014}.  
Specifically, WDM with a thermal mass of $2$ keV has been shown to be sufficient to 
solve some of the problems \citep{Anderhalden2013}. Although this mass is in conflict
with existing limits on free-streaming cutoffs \citep[e.g.,][]{Polisensky2011,Viel2013,Schneider2014}, 
the limits are subject to systematic uncertainties, and more robust limits based on 
phase-space arguments and subhalo counting are just below $2$ keV \citep{Boyarsky2008ju,Gorbunov2008ka,Horiuchi2014}.

The \bicep\ measurement may also have interesting consequences on searches 
for potential annihilation signals from dark matter itself (indirect detection
studies).  The annihilation signal from a single halo scales as the square of the
dark matter density, $\rho_{\rm DM}^2$ \citep{Strigari2008}, and the 
total ``boost" factor, the contribution to the expected annihilation signal 
due to substructure, is dependent on the slope and normalization of the 
substructure mass function.  Reducing any of these quantities could 
significantly loosen the upper limits placed by the searches that employ 
substructure boost \citep{Kamionkowski2010,Anderson2010,Sanchez-Conde:2013yxa,Ng2013}.

In this paper, we investigate the impact of the running power spectrum on
structure formation in the Universe by simulating the evolution of a MW-size
host in four separate cosmologies:  the model motivated by \bicep, the \planck\ 
cosmological model, a WDM model with the \planck\ parameter set, and a 
flat universe with a lowered $\om$ but otherwise identical to the \planck\
universe in order to control for the difference in $\om$ between the \planck\
and \bicep\ models. 

This paper is organized as follows:  \S~\ref{sec:sims} describes the simulations, 
including the cosmological models that we compare; \S~\ref{sec:results} presents 
our results for the cosmological mass function at $z = 3$, the subhalo $\vmax$ 
function of a MW-size host at $z = 0$, and discuss the changes in the internal 
kinematics of the highest mass subhalos (the TBTF problem) as well as implications
for the substructure boost; we summarize our findings in \S~\ref{sec:conclusions}.

\section{Simulations and Analysis}
\label{sec:sims}
We have run collisionless, dark matter-only 
simulations of a $50\hmpc$ periodic region with the Tree-PM code 
\texttt{Gadget-3} \citep{Springel2005}, beginning at $z = 125$.  
We present seven simulations, three of which model the full volume at 
medium resolution ($n_{\rm p} = 1024^3$) and four of which are ``zoom-in" 
simulations aimed at a Milky Way (MW)-size host.  Initial conditions were 
created with \texttt{MUSIC} \citep{MUSIC}.
We include the running in the \bicep\ universe by defining 
\begin{equation}
T'^2(k) = \left(\frac{k}{k_\star}\right)^{\frac{1}{2}\alpha_{\rm s} \ln\left(\frac{k}{k_\star}\right)}T^2(k),
\end{equation}
where $\alpha_{\rm s} = dn_{\rm s}/d\ln k$ is the running of the spectral 
index, $k_\star = 0.05\,{\rm Mpc}^{-1}$ \citep{KnottedSky2}, and $T(k)$ 
is the standard definition of the transfer function.  We pass $T'^(k)$ to 
\texttt{MUSIC} as the transfer function. 

\begin{table}
\centering
\begin{tabular}{lcccc}
Parameter   		     & \bicep & \planck & \litep & \wdmp \\ \hline \hline
$\alpha_{\rm s}$     & -0.024  & 0	   & 0	& 0	     \\
$h$				    		  & 0.698	  & 0.6711 & 0.6711	& 0.6711	 \\ 
$\om$ 				      & 0.285	  & 0.3175 & 0.26 & 0.3175	 \\
$\ol$ 						& 0.715	  & 0.6825 & 0.74 & 0.6825	 \\
$\sigma_8$				& 0.835	  & 0.8344 & 0.8344 &   0.8344	 \\
$n_{\rm s}$				& 0.967	  & 0.9624 & 0.9624 &   0.9624	 \\
$m_{\rm WDM}$	& --- & --- & --- & 2.6~keV \\
$m_{\rm p,\,HR}$  & 1.44  & 1.6 & 1.31  & 1.6 \\
$m_{\rm p,\,FB}$ & 92.1 & 102.6 & 84 & --- \\ 
\hline
\end{tabular}
\caption{The four sets of cosmological parameters used in this work.
  The first column indicates the parameter, the second lists the
  adopted \bicep\ cosmology from \citet{KnottedSky2}, the third gives
  the parameters from \planck\ adopted here \citep[taken from the
    temperature power spectrum;][]{PlanckCosmo}, the fourth column
  lists the ``\litep" cosmology, which is identical to the
  \planck\ parameter set but with an overall matter density below that
  suggested by \bicep.  The final column, which we refer to as
  ``\wdmp," is identical to the \planck\ cosmology, but includes a WDM
  free-streaming cut-off in the power spectrum for a thermal WDM
  particle mass of $m_{\rm WDM} = 2.6\rm keV$ (see
  Figure~\ref{fig:powerspec}).  Particle masses are given in units of
  $10^5\hmsun$.  $\alpha_{\rm s}$ is the running, defined in the
  text.}
\label{tab:cosmo}
\end{table}

We list the four underlying cosmological models that we adopt in
Table~\ref{tab:cosmo}.  For the \bicep\ universe, we select the
``running" model from \citet{KnottedSky2}, who performed a joint
Bayesian analysis on the \bicep\ $B$-mode polarization data and the
temperature and lensing data from \citet{PlanckCosmo}; those
parameters are listed in the first column.  We elect to compare this
model to that suggested by the \planck\ temperature power spectrum
data alone \citep[Table 2, Column 2 of][]{PlanckCosmo}, reproduced in
the second column.  We additionally simulate structure formation in
two \planck-like control models, \litep\ and \wdmp.  Both adopt the
majority of the \planck\ parameters, but \litep\ artificially lowers
the overall matter density, $\om$, to $\sim3\sigma$ below that
suggested by \citet{KnottedSky2} (while maintaining flatness) in order
to control for the lowered $\om$ in the \bicep\ cosmology.  The
\wdmp\ cosmology is identical to the \planck\ model, but imposes a
relativistic free-streaming cut-off in the power spectrum for a
thermal WDM particle equivalent mass of $m_{\rm WDM} = 2.6$~keV.
The mass is chosen to obey the robust limits from phase-space arguments
of MW dSphs galaxies and strict counting of M31 satellites \citep{Horiuchi2014},
and is also marginally consistent with measurements of the Ly-$\alpha$ 
at $3\sigma$ \citep{Viel2013}.  A WDM particle mass of $2$~keV
has been shown to solve small-scale issues in CDM \citep{Anderhalden2013},
but we opt for a slightly more massive particle in order to explore a value
distinct from other works.

\begin{figure}  
\centering
\includegraphics[width=.49\textwidth]{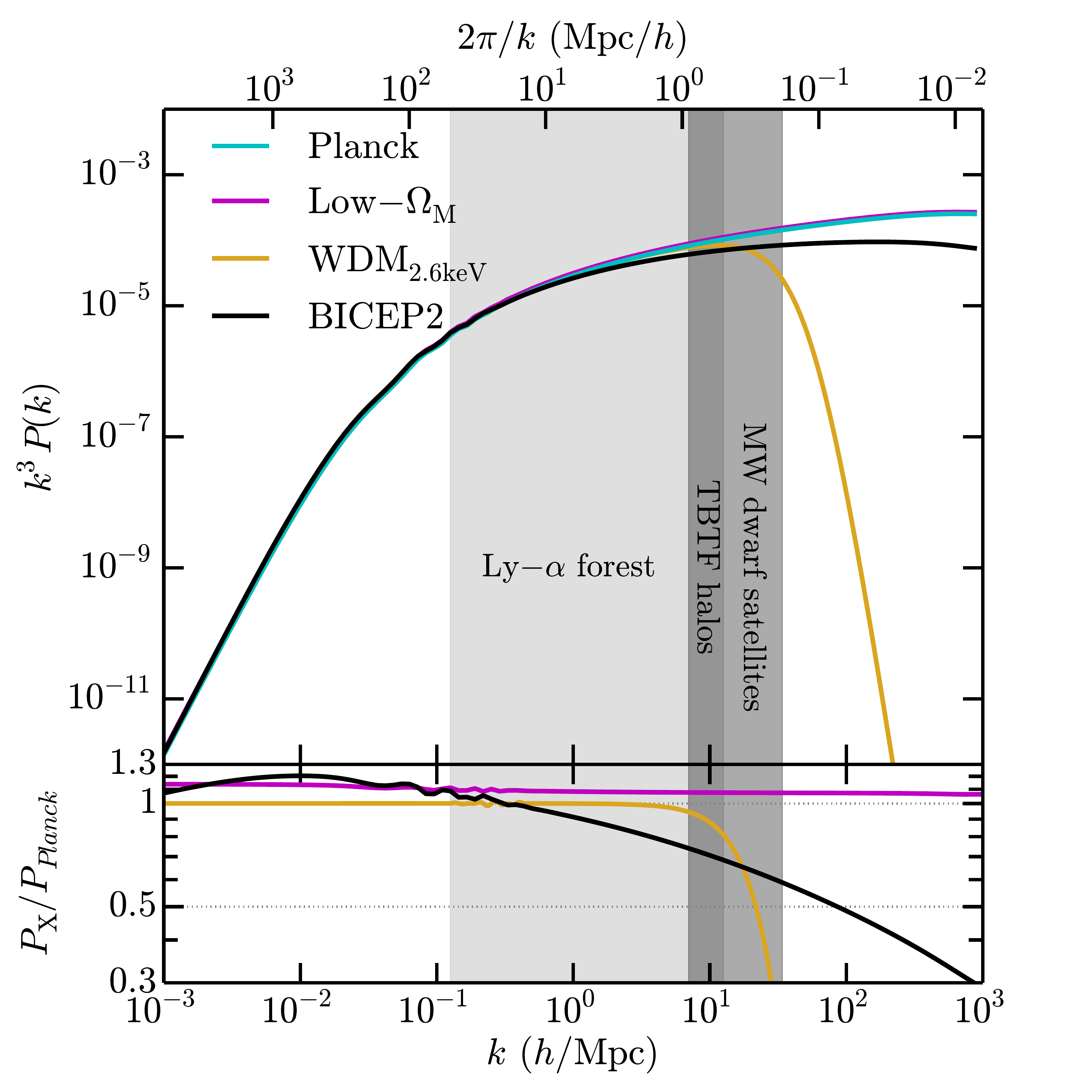}
\caption{Top:  The primordial power spectrum in the \bicep\ (black), \planck\ (cyan),
\wdmp\ (yellow), and \litep\ (magenta) cosmologies adopted in this paper, used 
for creating the initial conditions for the simulations.  Bottom:  The ratio of the power 
spectra relative to that of \planck.   The light shaded region in both panels 
indicates the regime that \citet{Viel2013} probe with the Lyman-$\alpha$ forest, 
where the \bicep\ power spectrum differs by $\lesssim30\%$ and where that of 
\wdmp\ agrees nearly perfectly, until the sharp cutoff just below the smallest scales
probed by Ly-$\alpha$.  On the mass scales relevant to small-scale galaxy formation 
($M_{\rm halo} \sim10^9-10^{11}\hmsun$, indicated in dark grey) however, \bicep\ 
differs by nearly a factor of $2$ and \wdmp\ quickly falls off due to relativistic 
free-streaming in the early Universe.  The overlap region roughly corresponds to
the mass scales of halos characteristic of the too-big-to-fail problem.  The \litep\ 
model is everywhere $\sim10\%$ higher than the standard \planck\ model at $z = 125$ 
due to the constraint that the linear power spectra agree at $z = 0$.}
\label{fig:powerspec}
\end{figure}

The initial ($z = 125$) matter power spectra for these cosmologies are
shown in Figure~\ref{fig:powerspec}.  The upper panel plots $k^3P(k)$
for the \bicep\ parameters in black, the \planck\ model in cyan, and
the \litep\ and \wdmp\ control models in magenta and yellow,
respectively.  The ratio of each model, relative to the \planck\ power
spectrum is plotted in the lower panel.  The light-grey region
indicates the scales that are currently probed by the Lyman-$\alpha$
forest \citep[$50\hmpc-0.5\hmpc$;][]{Viel2013} and the dark grey
region indicates the mass ranges of interest to dwarf galaxy formation
($M_\mathrm{halo}\sim10^9-10^{11}\msun$); the darkest overlap region
roughly corresponds to the mass scales of $\vmax\sim35~\kms$ halos,
which are characteristic of the problematic halos identified in TBTF.
The \bicep\ power spectrum differs from that of \planck\ by as much as 
$\sim30\%$ at the scales probed by the Ly-$\alpha$ forest; studies of 
the Ly-$\alpha$ forest power spectrum are sensitive to running, and the 
most recent results have found values consistent with the running we 
adopt here $\alpha_s = -0.028\pm 0.018$ \citep{Lesgourgues:2007te}.  
The $\gtrsim 30\%$ reduction in the primordial power at the smaller 
scales associated with the formation of dwarf halos, however, has 
interesting consequences for the small-scale problems discussed above.  
The unlabeled region to the right of the dwarf scales are associated 
with so-called ``mini-halos," which may be probed by gravitational 
lensing studies \citep[e.g.,][]{Keeton:2008gq} or tidal stream analyses
\citep{Ngan:2013oga}.  This range is also important for the overall
``boost" factor due to dark matter annihilation in substructure
\citep{Sanchez-Conde:2013yxa}, indicating that the \bicep\ power
spectrum will likely produce a much smaller DM annihilation signal
from these mini-halos.

\begin{figure}  
\centering
\includegraphics[width=\columnwidth]{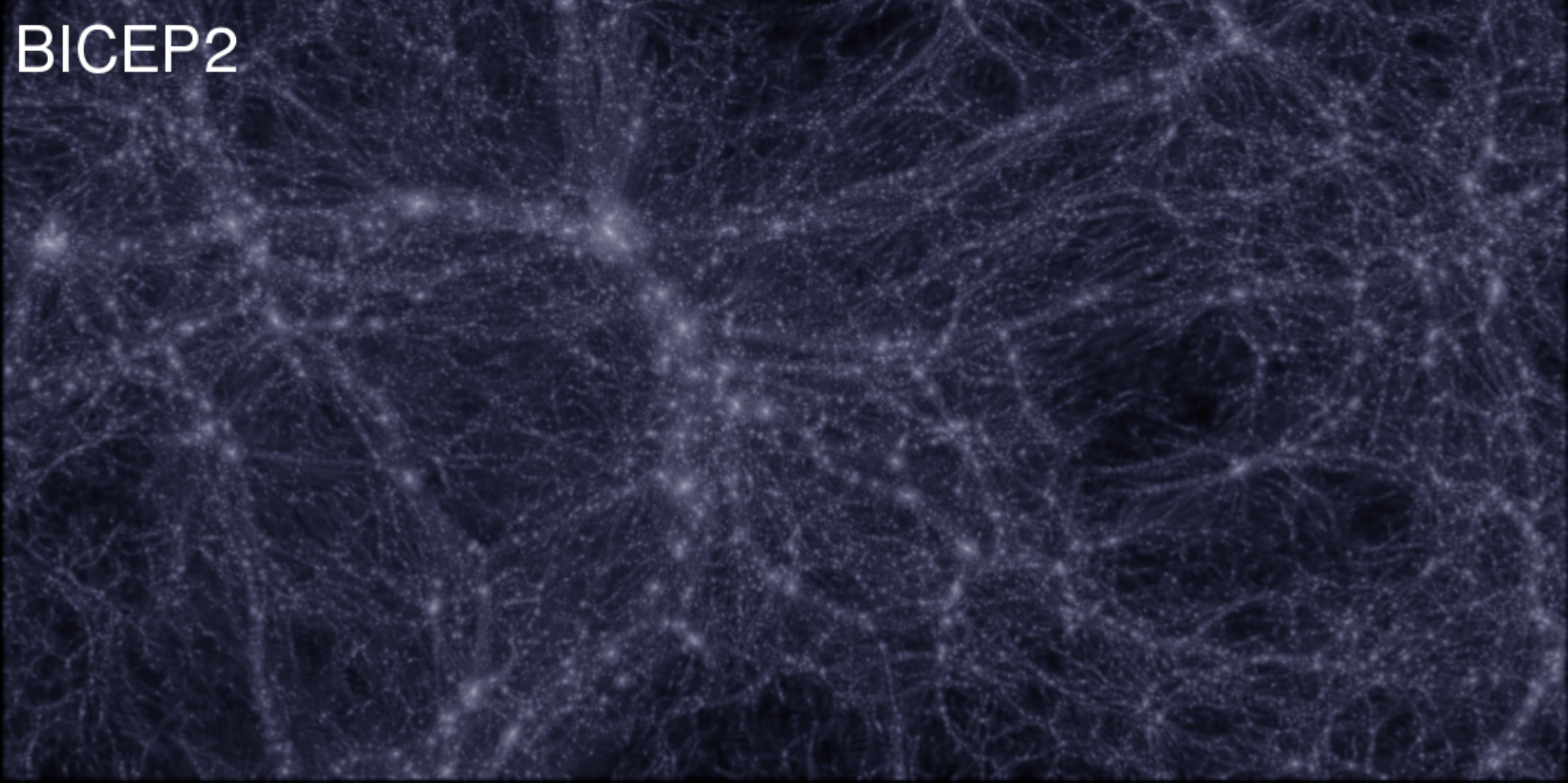}
\includegraphics[width=\columnwidth]{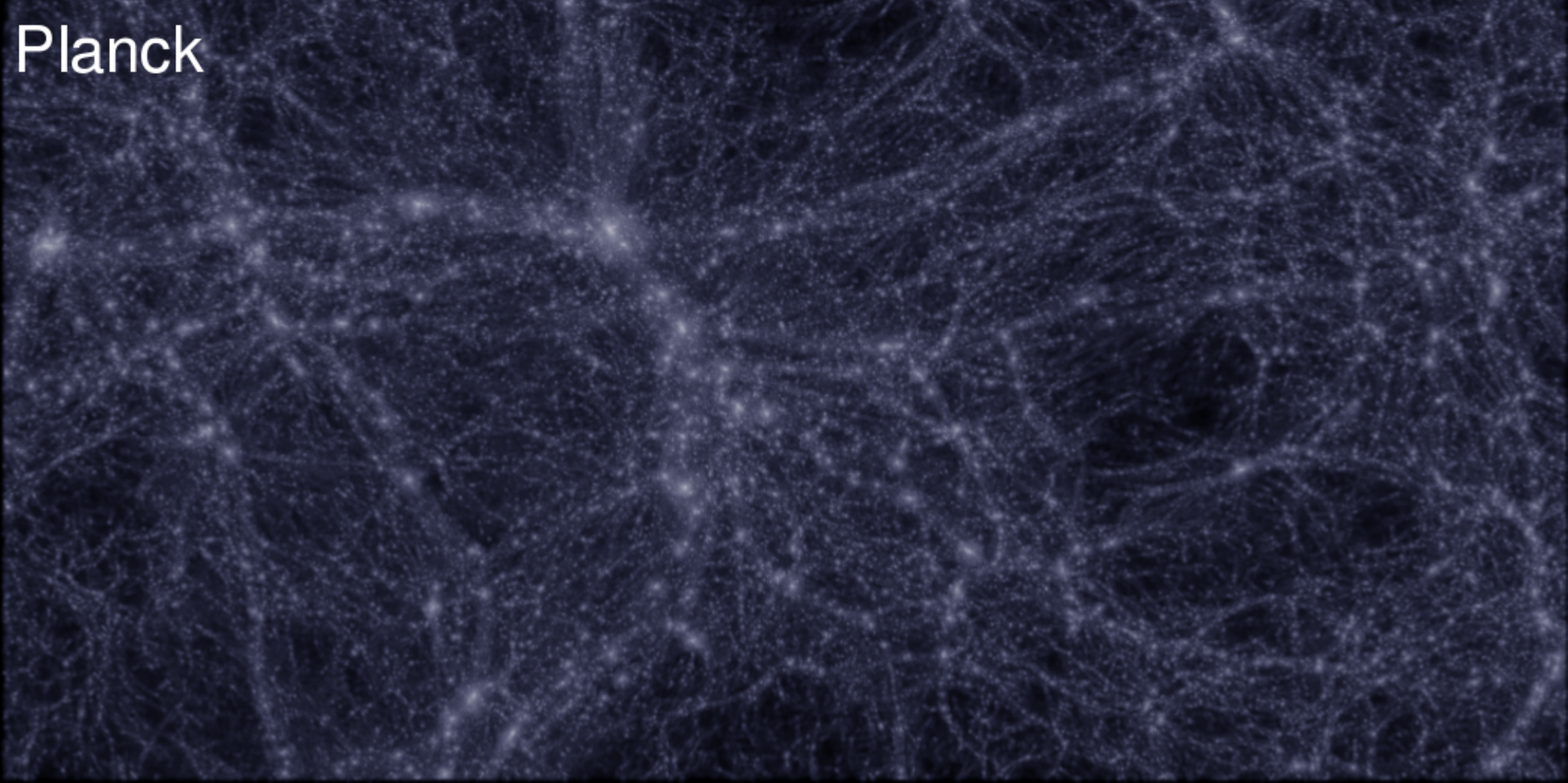}
\caption{Visualizations of the large-scale dark matter density field at $z = 3$ in
the \bicep\ (top) and \planck\ (bottom) cosmologies.  Shown is a slab $25~\hmpc$
wide, $12.5~\hmpc$ tall, and $5~\hmpc$ deep.  The two matter fields initially 
appear indistinguishable on these scales, though we will show below that there 
are small differences in the halo mass function, which become even stronger on 
the scales of dwarf galaxies. }
\label{fig:fbviz}
\end{figure}

We first compare the cosmologies by simulating the entire $50~\hmpc$ volume
at moderate resolution ($n_\mathrm{p} = 1024^3$) until $z = 3$ with the 
\planck, \litep, and \bicep\ cosmologies.\footnote{We do not simulate the 
   volume with the \wdmp\ cosmology as the model is designed to agree 
   with our \planck\ run at the scales probed by such a simulation.}  
The particle masses for these ``full-box" simulations are given in 
Table~\ref{tab:cosmo} as $m_{\rm p,\,FB}$ in units of $10^5\msun$.  We 
fix the Plummer-equivalent softening lengths of the full-box simulations at 
$5$~comoving~$\hkpc$ until $z = 9$, at which time they become 
$500$~physical~$h^{-1}$~pc.  Dark matter structure is identified with 
the \texttt{AMIGA Halo Finder} \citep[\texttt{AHF};][]{AHF}, a publicly-available 
three-dimensional spherical overdensity halo finder.\footnote{\texttt{AHF} is 
available at \url{http://popia.ft.uam.es/AHF/Download.html}.}  A slice of the 
simulation volume at $z = 3$ is shown in Figure~\ref{fig:fbviz} for the \bicep\ 
cosmology (top) and the fiducial \planck\ model (bottom)~--~the two appear 
indistinguishable at these scales, though we will show below that there is a small 
systematic offset in the halo mass function, consistent with expectations from 
linear theory.

\begin{figure} 
\centering
\includegraphics[width=0.495\columnwidth]{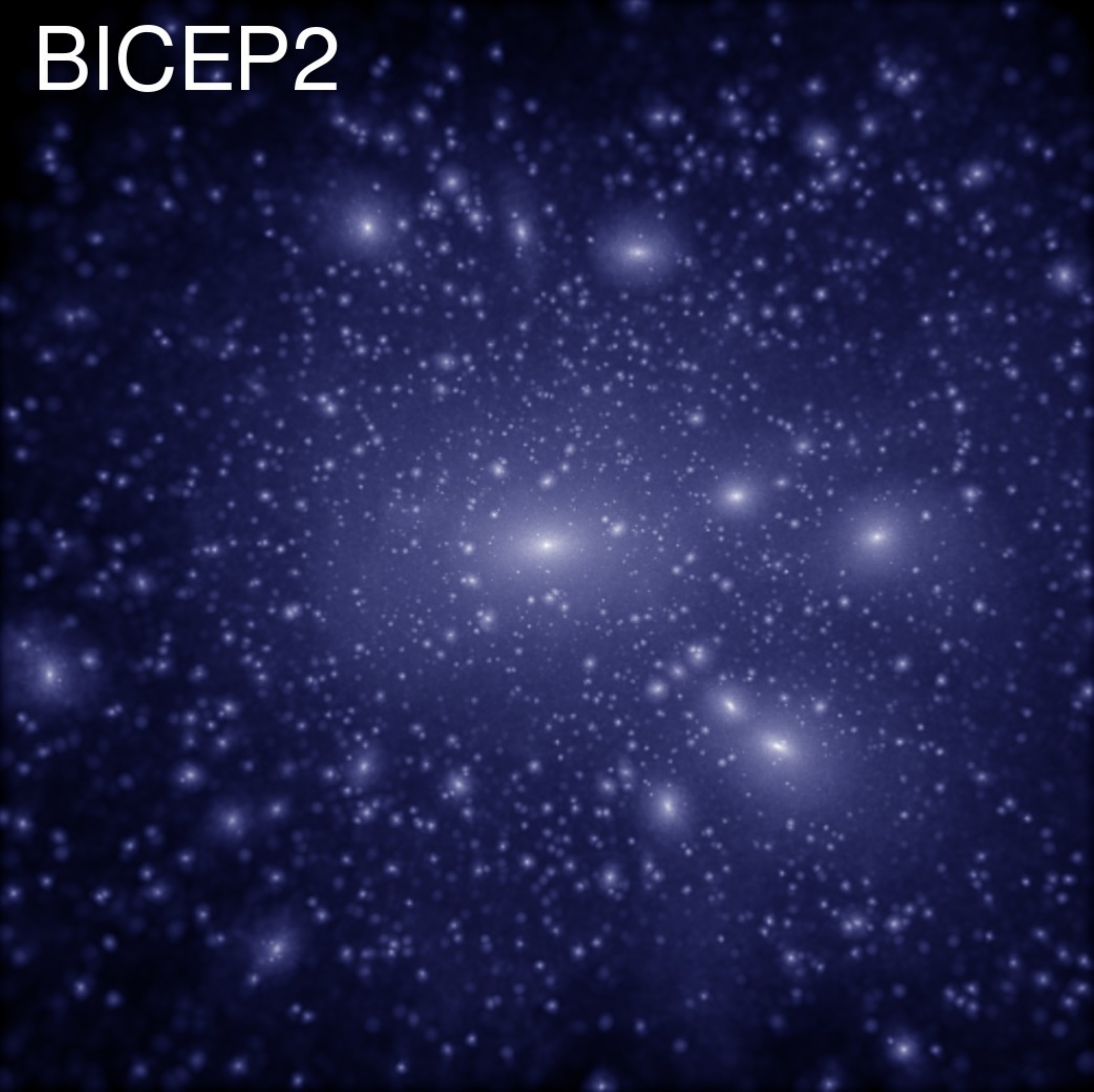} \hspace*{-0.5em}
\includegraphics[width=0.495\columnwidth]{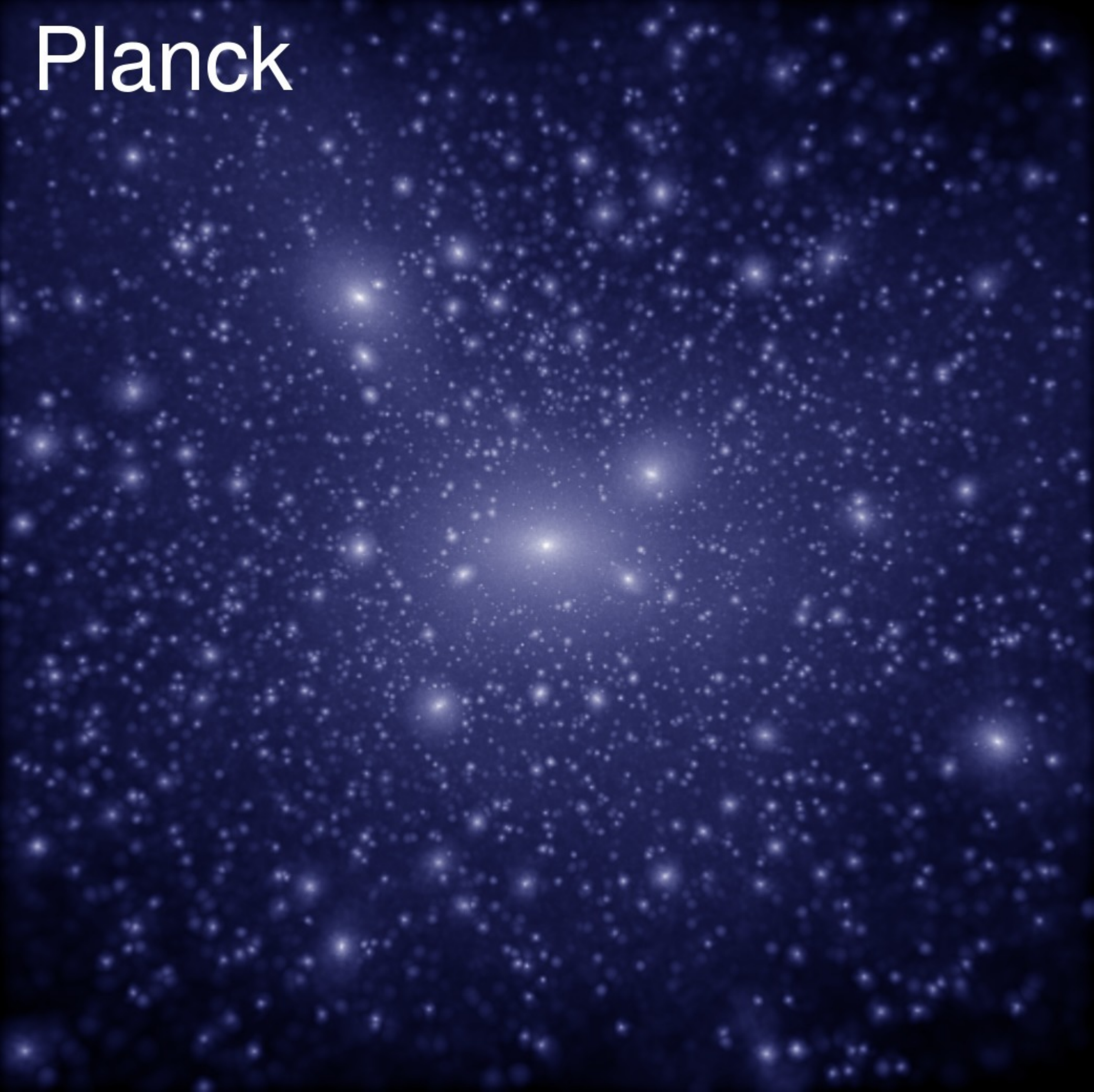} 
\includegraphics[width=0.495\columnwidth]{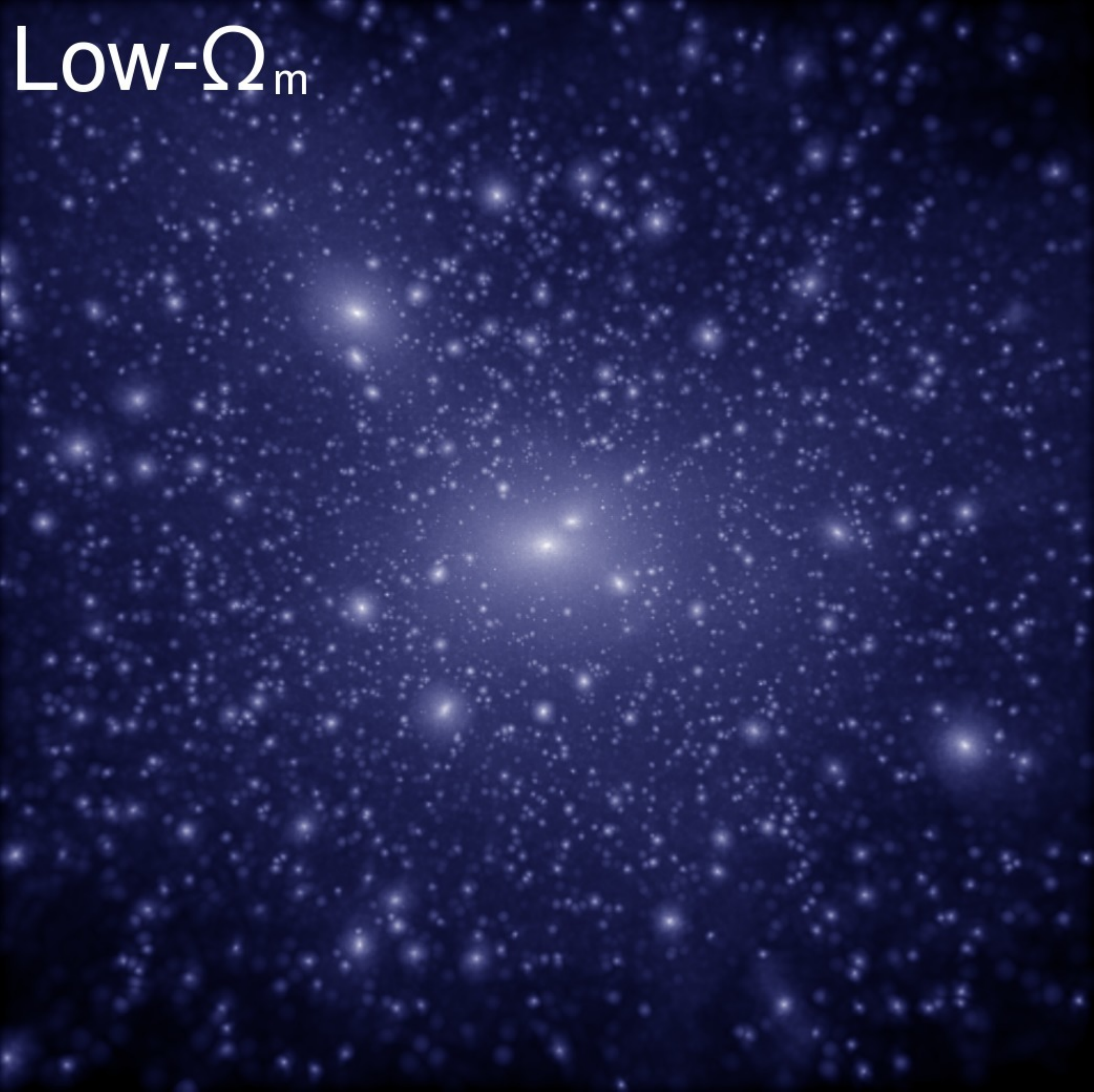} \hspace*{-0.5em}
\includegraphics[width=0.495\columnwidth]{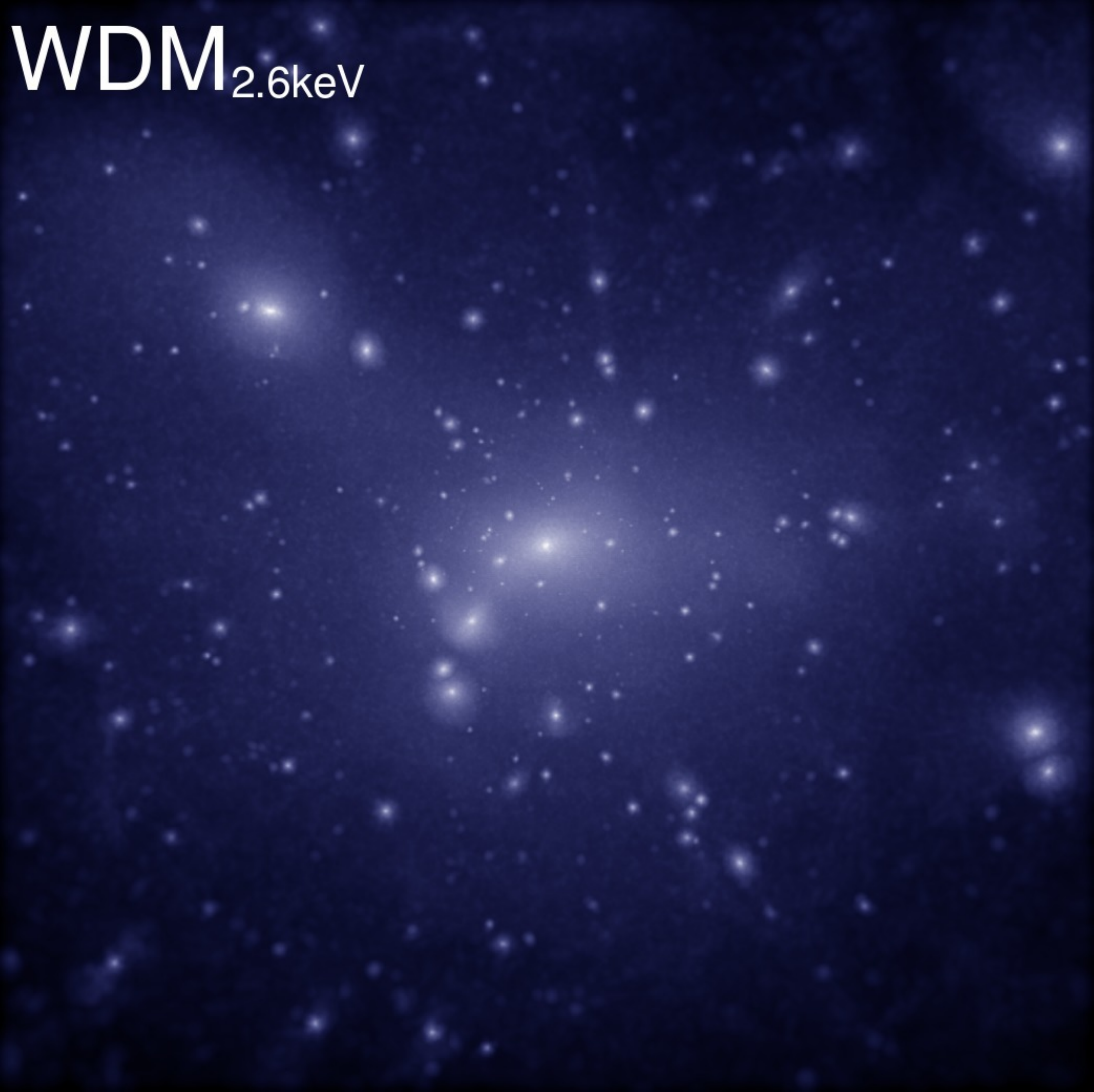}
\caption{Visualizations of the zoom-in halo, colored by the local matter 
density, in the \bicep\ (top left), \planck\ (top right), \litep\ (bottom left), 
and \wdmp\ (bottom right) cosmologies.  Shown are cubes $500~\hkpc$ on a side, 
centered on the targeted host.  The relative lack of substructure in the \wdmp\ run
and the agreement between the orbits of the largest halos between the \planck\ 
models are visible even by eye.  The \bicep\ simulation, however, displays clearly
different subhalo orbits and hints at reduced substructure at the smallest masses.}
\label{fig:zoomviz}
\end{figure}

In order to study the highly non-linear regime, however, we primarily focus 
our efforts on ``zoom-in" simulations \citep{Katz1993,Onorbe2013} aimed at 
a MW-size host, similar to the Via Lactea II \citep{VL2Nature,VL2APJ} and
Aquarius \citep{Aquarius} projects.  Specifically, we selected a highly 
isolated host from the ELVIS simulations \citep{ELVIS} and re-create the 
parent box, oversampling the region from which the halo forms with higher 
resolution, with the four underlying cosmological models given in Table~\ref{tab:cosmo}.
The zoom-in simulations are initialized with an effective resolution of 
$4096^3$ particles in the high resolution region.  Similar to the full-box 
simulations, the softening lengths of these lowest mass particles is kept 
fixed at $1$~comoving~$\hkpc$ until $z = 9$, after which it is held fixed 
at $100$~physical~$h^{-1}$~pc until $z = 0$.  The particle masses for each 
cosmological model are listed as $m_{\rm p,\,HR}$ in Table~\ref{tab:cosmo},
again in units of $10^5\hmsun$.  Each cosmological model was initialized 
with identical phases for the perturbations at all scales in order to reduce
numerical differences (e.g., in the subhalo orbits) between the models.
As in the full-box simulations, we search for collapsed structures in the
$z = 0$ particle data with \texttt{AHF}.\footnote{
	We also find identical results using the 6D friend-of-friends 
	halo finder \texttt{ROCKSTAR} \citep{rockstar}.}
A visualization of a cube $500~\hkpc$ on a side, centered on the zoom-in 
target, is shown in Figure~\ref{fig:zoomviz}.  The images are colored by the 
local matter density and show, from top left to bottom right, the \bicep\ 
simulation, the \planck\ model, the \litep\ cosmology, and the \wdmp\ 
model.  The agreement between the \planck\ models, in spite of the 
free-streaming cutoff or shift in $\om$, is uncanny; the \bicep\ cosmology, 
however, has less overall substructure and clearly distinct orbits for the 
largest subhalos, indicative of the significant differences in power at 
$M \sim 10^{9} - 10^{11} \msun$ scales seen in Figure~\ref{fig:powerspec}.

Our zoom-in simulations are run with identical particle numbers, box
sizes, and softening lengths (in $h^{-1}$ units) as the fiducial
simulations in the ELVIS Suite \citep{ELVIS}; we therefore adopt the
ELVIS resolution cut here and study only halos with maximum circular
velocities $\vmax > 8~\kms$.  Similarly, \citet{ELVISTBTF} showed that
the relationship between $\vmax$ and the radius at which $\vmax$
occurs, $\rmax$, is converged for halos larger than $15~\kms$ and with
$\rmax > 0.36~\hkpc$ for simulations at this resolution; we again use
the same criteria when examining the internal structure of small
halos.

\section{Results}
\label{sec:results}
\begin{figure}  
\centering
\includegraphics[width=\columnwidth]{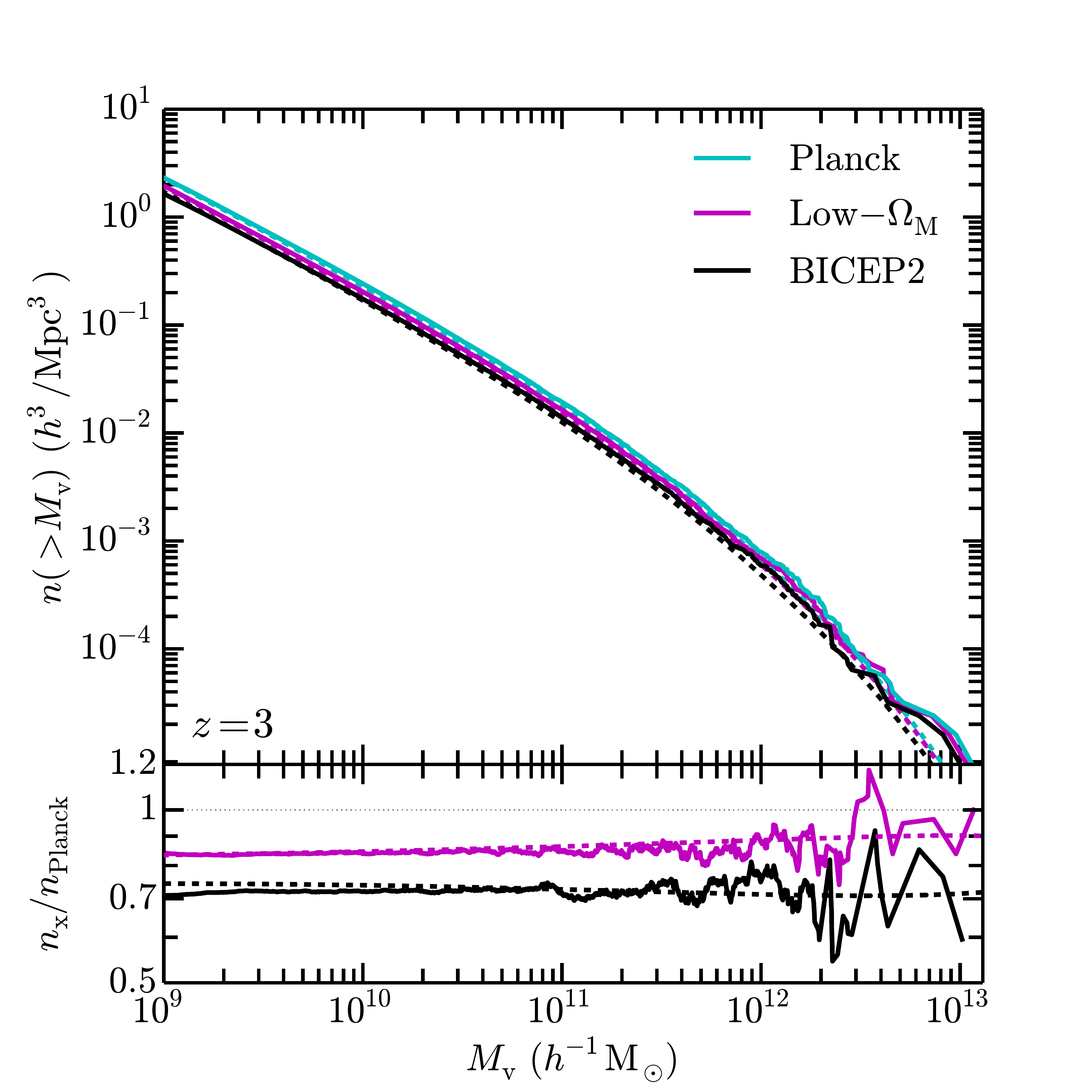}
\caption{The anti-cumulative mass function, per unit volume, of all host halos in 
the $50~\hmpc $ volume at $z = 3$ from the simulations (solid lines) and from
applying the \citet{Tinker2008} analytical fitting function (dashed lines) for the 
\bicep\ (black), \planck\ (cyan) and \litep\ (magenta) models (upper panel)
and the ratios of the \bicep\ and \litep\ models to the \planck\ model (lower 
panel).  At fixed mass, the \bicep\ cosmology predicts $\sim30\%$ fewer halos
than the \planck\ model, consistent with expectations from linear theory.
Alternatively, halo masses at fixed number counts are $\sim20-30\%$ lower
in the \bicep\ model, again compared to \planck.}
\label{fig:fbmassfunc}
\end{figure}

We begin by examining the halo mass function in the $50~\hmpc $ full-box
runs at $z = 3$.  Plotted as solid lines in the top panel of Figure~\ref{fig:fbmassfunc} 
is the anti-cumulative number density of host halos, defined as those with their centers 
outside the virial volumes
	\footnote{We use the term ``virial radius" to refer to the 
	radius at which the overdensity relative to the critical density 
	drops to 173.8 (\bicep), 174.3 (\planck), and 173.3 (\litep) at 
	$z = 3$ and 99.8 (\bicep), 104.1 (\planck), 96.5 (\litep), and 
	104.1 (\wdmp) at $z = 0$, and ``virial mass" to refer to the total 
	mass contained within that radius.}
of all halos larger than itself, as a function of virial mass $\mvir$; the 
lower panel plots the ratio of each line relative to the fiducial \planck\ model.  
The \bicep\ cosmology exhibits a suppression on all mass scales such that the 
Planck mass function is offset by $\sim30\%$ at fixed number, though the offset 
rises slighter at lower masses, consistent with the running in the power 
spectrum.  We note that presenting results in $\msun$ rather than $\hmsun$ 
would only increase the difference between the two simulations as the \planck\
cosmology adopts a smaller Hubble parameter.

This offset, however, is consistent with expectations from linear
theory of structure collapse.  Plotted as dashed lines in
Figure~\ref{fig:fbmassfunc} are the results of applying the analytical
fitting function of \citet{Tinker2008};
  \footnote{Theoretical mass functions are
  calculated via the publicly available code provided by
  \citet{HMFcalc}.} 
the ratios of these fitting functions are plotted as dashed lines in 
the lower panel.  The \citeauthor{Tinker2008} fit agrees nearly perfectly 
with our simulated mass functions, and the relative offsets from the 
\planck\ model are also in excellent agreement with the simulations.  
We conclude that analytic mass functions based on linear theory may 
be used to make accurate predictions (at least until $z=3$) in the 
\bicep\ cosmology.

Given that the differences in the primordial power spectrum increase 
with decreasing scales, we can expect to see even more extreme 
differences on the scales of dwarf galaxy halos.  We therefore turn 
our analysis to the zoom-in simulations described in Section \ref{sec:sims},
which we exclusively use for the remainder of the work.  The properties 
of the main host halo, given in Table~\ref{tab:props}, vary slightly 
between the four models; we therefore present subhalo counts as a 
function of $\vmax/\vvir$, where $\vvir$ is the circular velocity of the 
host halo at the virial radius.  This minimizes the halo-to-halo scatter 
and normalizes for the effects of varying host mass. 

\begin{table*}
\centering
\begin{tabular}{lcccccccc}
				 & $\mvir$ & $\rvir$ & $\vmax$ & $\vvir$ & $N_{\rm h}(<\rvir)$ & $N_{\rm p}(<\rvir)$ & $r_{\rm uncontam}$ & $N_{\rm h}(<1\,\hmpc)$\\ 
				&	($10^{12}~\hmsun$) & ($\hkpc)$ & $(\kms)$ & $(\kms)$ & [$>8~\kms$]  &  & ($\hmpc$) & [$>15~\kms$] 
\\ \hline \hline
\bicep & 1.26 & 221 & 164 & 156 & 460 & 8.8$\times10^6$ & 1.27 & 125 \\  
\planck & 1.49 & 231 & 187 & 166 & 944 & 9.4$\times10^6$ & 1.04 & 216 \\ 
\litep & 1.21 & 222 & 176 & 153 & 709 & 9.3$\times10^6$ & 1.05 & 166 \\ 
\wdmp & 1.49 & 231 & 194 & 167 & 119 & 9.4$\times10^6$ & 0.97 & 76 \\ 
\hline
\end{tabular}
\caption{The properties of the main host halo in the zoom-in simulations.  In order, 
the columns are the virial mass $\mvir$, virial radius $\rvir$, maximum circular 
velocity $\vmax$, virial velocity $\vvir = \sqrt{G\mvir/\rvir}$, the number of 
resolved ($\vmax > 8~\kms$) subhalos within the virial radius $N_{\rm h}(<\rvir)$, 
the number of simulation particles within the virial radius $N_{\rm p}$, the 
distance to the nearest low resolution particle $r_{\rm uncontam}$, and the number 
of halos with resolved internal structure ($\vmax > 15~\kms$) within $1~\hmpc$, 
$N_{\rm h}(<1~\hmpc)$.}
\label{tab:props}
\end{table*}

\begin{figure}
\centering
\includegraphics[width=\columnwidth]{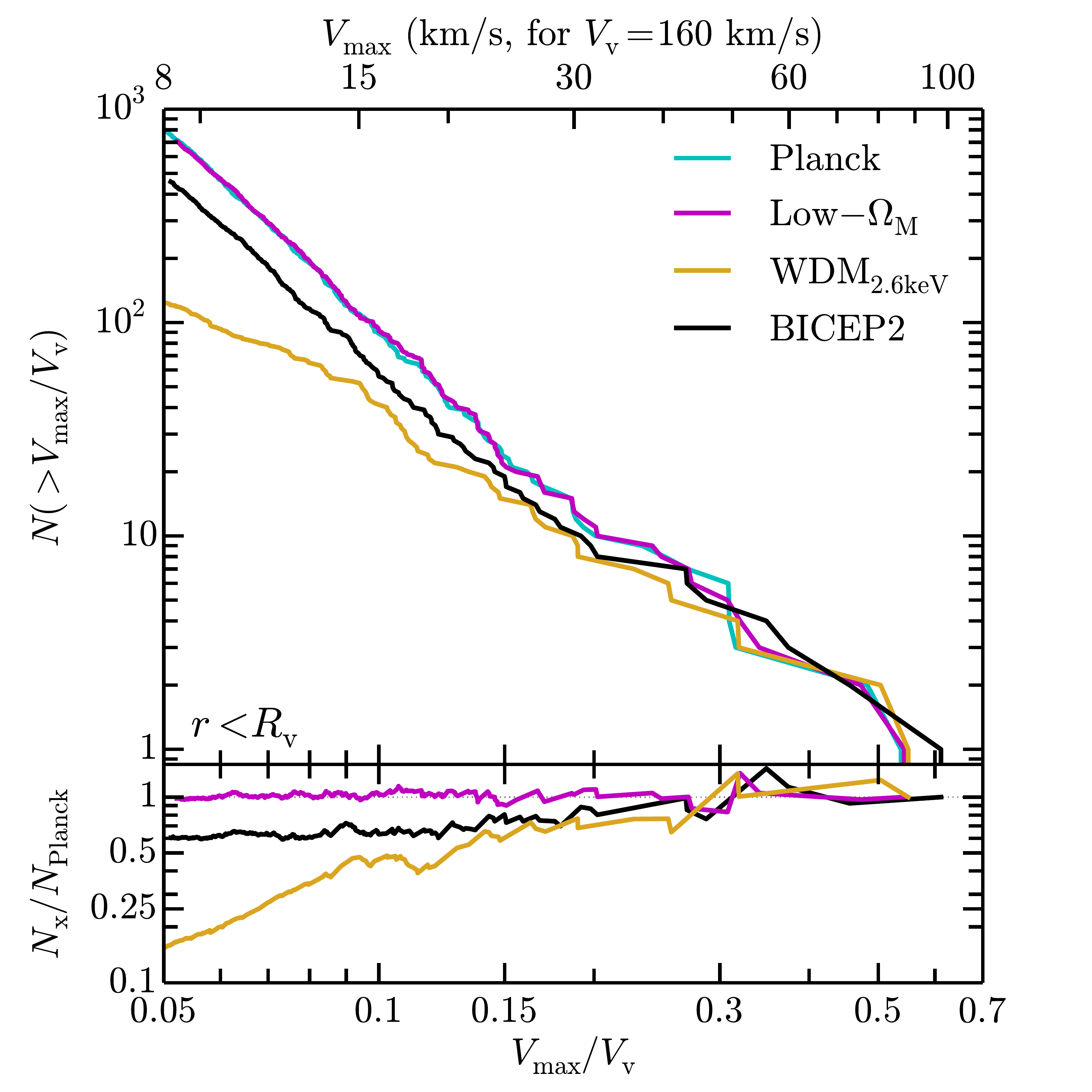}
\caption{The anti-cumulative count of subhalos ($r < \rvir$) as a function 
of $\vmax$ normalized by the host virial velocity, $\vvir$ (upper panel) and
the ratio of each cosmology to the \planck\ model (lower panel). Counts in 
the \litep\ cosmology (magenta line) match up nearly exactly with those in the 
standard \planck\ cosmology (cyan line), even though the host halo is $\sim20\%$ 
less massive due to the modification in $\om$.  Counts in the \bicep\ cosmology 
(black line), however, are systematically low for $\vmax/\vvir \lesssim 0.25$ 
($\vmax \lesssim 40~\kms$) and predict $\sim50\%$ fewer halos at the resolution 
limit.  The \wdmp\ model, meanwhile, drastically under-produces subhalos at low 
masses.  Therefore, both the \wdmp\ and the \bicep\ model will alleviate the 
missing satellites problem, though \wdmp\ may eliminate too many subhalos to 
explain, e.g., the observed satellite mass function of M31 \citep{Horiuchi2014}.  
The top axis is scaled to $\vvir = 160~\kms$, the mean virial velocity of the 
host in the four simulations.}
\label{fig:vmaxfunc}
\end{figure}

This normalized $\vmax$ function is plotted in the top panel of 
Figure~\ref{fig:vmaxfunc} for all four cosmological models; the lower
panel again plots the ratio of each model to the \planck\ cosmology.  The 
upper axis is scaled to $\vvir = 160~\kms$, roughly the virial velocity of a 
MW-size host and the mean $\vvir$ of the host in the four simulations.  
When normalizing by $\vvir$, the agreement between the \planck\ and \litep\ 
models is nearly perfect at all $\vmax/\vvir$, even at the high $\vmax$ 
end where small-number statistics typically dominate; if the counts are not 
normalized by the virial velocity, however, the \litep\ model lies $\sim25\%$ 
below the Planck cosmology at fixed subhalo $\vmax$.  The \bicep\ counts, 
however, are suppressed even after normalizing by $\vvir$, particularly for 
subhalos less massive than $\vmax\sim30~\kms$.  The total count is $\sim 50\%$ 
below the \planck\ line at the resolution limit, alleviating the severity of 
the missing satellites problem.  As expected, subhalos are even more strongly 
suppressed in the \wdmp\ universe, with counts a factor of $\sim6$ lower 
than the fiducial \planck\ model at the resolution limit. While this 
suppression drastically reduces the severity of the missing satellites 
problem, it may actually under-produce the required subhalo count compared 
to the known count of M31 satellite galaxies \citep[e.g.][]{Horiuchi2014}.\footnote{
	WDM N-body simulations are known to suffer from artificial fragmentation
	on small scales, leading to a non-negligible contribution to the halo
	catalog from spurious objects \citep[e.g.][]{Wang2007,Lovell2013}.  We
	do not explicitly account for this effect, which would act to suppress 
	counts of small halos even further.}  
The \bicep\ model has no such difficulties.

Due to the overall suppression of substructure in \bicep\, it is possible 
that counts of high mass ($\vmax \sim 80~\kms$) satellites will provide an
additional constraint on the running.  While we do not see any significant
differences in the few subhalos that exist in the simulated host at that 
mass range, it is possible that some reduction exists on a statistical 
level, particularly for close pairs.  As \citet{Tollerud2011} showed 
that \lcdm\-like cosmologies reproduce observations reasonably well at 
Large Magellanic Cloud (LMC)-like masses, such counts may be used as 
a probe of the initial power spectrum in the future.  Such a study, however,
would require large simulations with higher resolution than those presented 
here, simulated until $z = 0$.

We now turn our attention to the internal structure of the subhalos.  The 
simulations used in this work do not fully resolve density profiles in the
innermost $\sim500$~pc of dwarf halos, but integral properties such as $\vmax$ 
and $\rmax$ are converged for $\vmax>15~\kms$ objects.  These two quantities fully
define the two-parameter Navarro-Frenk-White \citep[NFW; ][]{NFW1996} density 
profile
\begin{equation}
\rho(r) = \frac{\rho_\mathrm{s}}{(r/r_\mathrm{s})(1+r/r_\mathrm{s})^2},
\label{eqn:nfw}
\end{equation}
where $r_{\rm s} = \rmax/2.1626$ is a characteristic scale radius and 
$\rho_{\rm s} = \rho_{\rm s}(\rmax,\vmax)$ is four times the density 
at $r = r_{\rm s}$.  We may therefore extrapolate a unique circular 
velocity curve into the inner regions of the halos to make predictions 
regarding the central densities and compare with observations at small 
radii.  This extrapolation assumes that the inner structure of subhalos 
is not strongly dependent on cosmology (i.e. that subhalos still follow 
NFW profiles in \bicep);  for \wdmp\ at least, this extrapolation seems to be valid 
\citep{Dunstan2011}, but we note that varying the density profile can 
strongly impact the number of massive failures \citep{diCintio2013,ELVISTBTF}.  
Similarly, we may predict the relative change in the annihilation signal 
from substructure by knowing only the relationship between $\vmax$ 
and $\rmax$, as the signal from a single halo or subhalo is proportional 
to $\rho_{\rm s}^2r_{\rm s}^3$ \citep{Strigari2008}.

\begin{figure}
\centering
\includegraphics[width=0.49\textwidth]{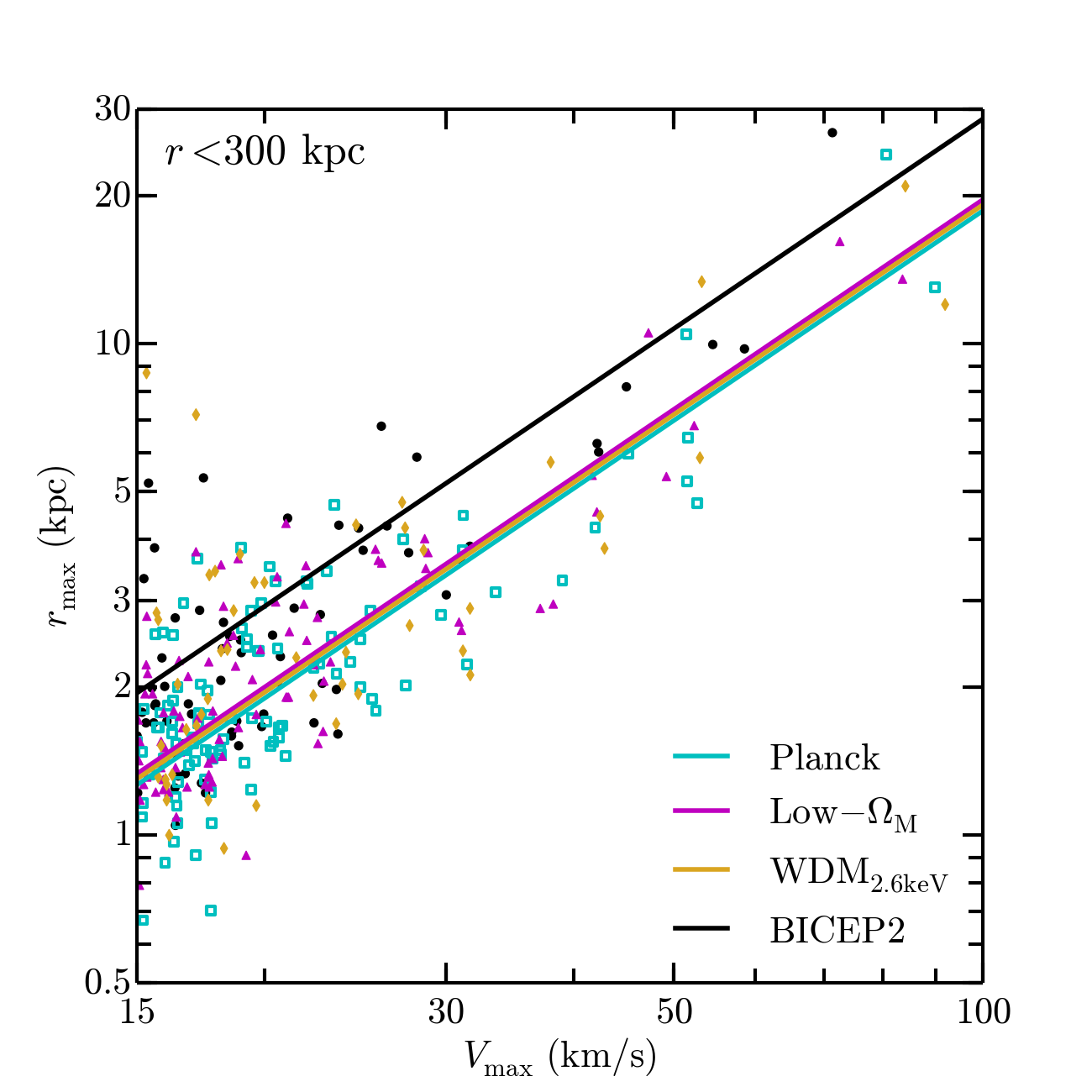}
\caption{The relationship with $\rmax$ and $\vmax$ for subhalos
in the \bicep\ (black circles), \planck\ (cyan squares), \litep\ 
(magenta triangles), and \wdmp\ (yellow diamonds) cosmologies, along
with power-law fits to the data (Equation~\ref{eqn:rmaxfit}).  
The fits are weighted by $\vmax$ with the log-slope held fixed at the best-fit 
value in the \planck\ model,  $p = 1.419$ (though there are weak indications 
that the slope is steeper in the \bicep\ model).  The best-fit normalization in 
the \bicep\ cosmology is 35\% lower than in the \planck\ simulation.  In addition 
to helping to alleviate TBTF (see Figure~\ref{fig:vcirc}), this overall shift in 
$\rmax$ at fixed $\vmax$ also implies a $\sim35\%$ lower annihilation signal from 
each subhalo in \bicep.  The normalizations, $A$, are 0.71 (\planck), 0.75 (\litep), 
0.73 (\wdmp), and 1.09 (\bicep).}
\label{fig:rmaxvmax}
\end{figure}

We thus begin our investigation by presenting this relationship
for subhalos of the main host (within $300$~physical~$\kpc$, for 
comparison to the MW satellites) in the four cosmological models.
Plotted in Figure~\ref{fig:rmaxvmax} are the individual $\rmax-\vmax$
values for subhalos in each model, with the \bicep\ model plotted as black circles,
the \planck\ model in cyan squares, the \litep\ model as magenta triangles,
and the \wdmp\ model as yellow diamonds.  The lines plot power-law fits 
to the subhalos:
\begin{equation}
\frac{\rmax}{1~{\rm kpc}} = A\left(\frac{\vmax}{10~\kms}\right)^p.
\label{eqn:rmaxfit}
\end{equation}
The contribution to the least-squares fit from each halo is weighted by the 
$\vmax$ of that halo to account for the scarcity of high $\vmax$ halos, and
the log-slope $p$ is held fixed at the value that best fits the data in the 
\planck\ cosmology, $p = 1.419$, allowing the normalization $A$ to vary.
	\footnote{We have also tested a quadratic fit in 
	log-space and do not find evidence for a roll-off at small $\vmax$, though
	there are weak indications that the slope is steeper for the \bicep\ subhalos.}  
The three \planck-like models agree nearly perfectly: the normalizations differ by 
only $5\%$.  The \bicep\ model, however, is clearly offset from the remaining 
three cosmologies with a normalization 35\% higher. 

It is interesting to note that the \wdmp\ model yields similar subhalo structural 
parameters ($\vmax$ - $\rmax$) to those of the \planck\ models, at least for the 
velocity range plotted here.  Below we show that this is {\em not} the case for field
halos in \wdmp, which are less concentrated than \planck\ halos in the field.
We interpret this differences as an effect of enhanced subhalo stripping for the 
\wdmp\ subhalos.  Host halos tend to strip matter from the outer parts of 
subhalos, making them more concentrated with time.  The \wdmp\ host halo 
density and mass remain similar to that in \planck\ cosmology, and  the relative 
stripping experienced by the low-concentration infalling subhalos is more significant 
than it is in any of the other models.  This is also consistent with the fact that we see 
many fewer subhalos in the \wdmp\ case.

\begin{figure*}
\centering
\includegraphics[width=\textwidth]{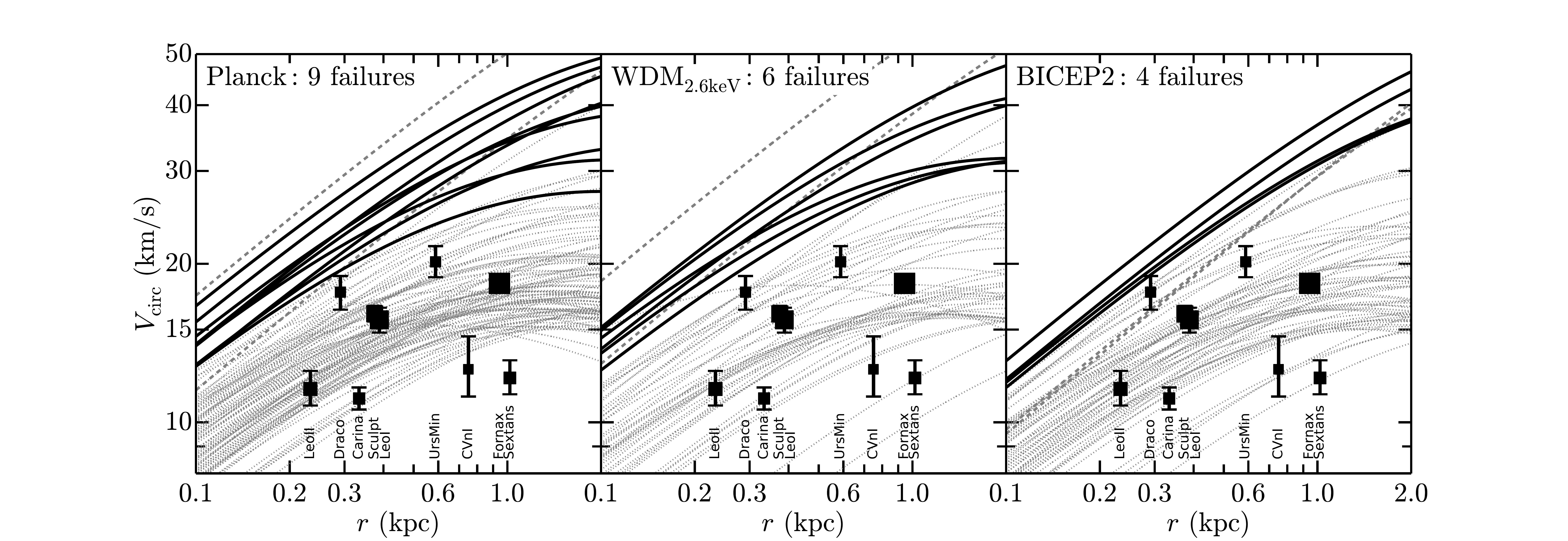}
\caption{The rotation curves of all halos within $300~\kpc$ of the host 
center with $\vmax > 15~\kms$, the smallest scale at which $\rmax$ can 
be reliably measured, in the \planck\ cosmology (left), the \wdmp\ model
(center), and the adopted \bicep\ cosmology (right).  The curves are 
extrapolated from $\rmax$ and $\vmax$ (Figure~\ref{fig:rmaxvmax})
by assuming an NFW profile.  Also plotted are the constraints on the circular 
velocity at the half-light radius of the nine classical MW dwarfs used to define 
the TBTF problem in \citet{MBK2011,MBK2012} from \citet{Wolf2010}.  
Plotted as solid lines are those halos identified as massive failures~--~subhhalos 
that lie above the $1\sigma$ constraints on the MW dwarfs and thus cannot 
host any of the known bright satellites.  As expected from 
Figure~\ref{fig:rmaxvmax}, which shows that subhalos in the \bicep\ cosmology 
are less dense at fixed $\vmax$ than in either the \planck\ or the \wdmp\ models, 
the problem is significantly alleviated (though not eliminated) by switching to the 
\bicep\ cosmology. For comparison, we note that the same halo contains 
eight massive failures in the \litep\ model.}
\label{fig:vcirc}
\end{figure*}

The differences seen in Figures~\ref{fig:vmaxfunc}~and~\ref{fig:rmaxvmax}
impact the counts of discrepant TBTF halos.  We directly compare the circular 
velocity curves predicted for each of our runs to observations of the classical 
MW dwarf spheroidal (dSphs) galaxies in Figure~\ref{fig:vcirc}~--~
each line represents a single subhalo within 300~kpc and each 
point indicates a MW satellite.  The left panel plots the \planck\ 
model, the central panel indicates the results in \wdmp, and
the right panel plots subhalos in the adopted \bicep\ cosmology.  
As in \citet{MBK2011,MBK2012}, the  observational sample is comprised 
of the galaxies within $300$~kpc of the MW with $L > 10 ^5~\lsun$, 
excluding the Magellanic Clouds and the Sagittarius dwarf.  The former 
is removed from the sample because satellites as large as the Clouds 
are rare around MW-size hosts \citep{Boylan-Kolchin2010,Busha2011,Tollerud2011}; 
we remove the latter because it is currently interacting with the MW disk 
and is therefore not in equilibrium.  For the remaining dwarfs, we plot 
$\vhalf$ at $\rhalf$, the circular velocity at the half-light radius, 
with $1\sigma$ errors in Figure~\ref{fig:vcirc}.  The values are taken 
from \citet{Wolf2010}, who used data from \citet{Walker2009}, \citet{Munoz2005}, 
\citet{Koch2007}, \citet{Simon2007} and \citet{Mateo2008}.

The lines in Figure~\ref{fig:vcirc} each indicate an NFW rotation 
curve for a single subhalo of the central host.  The dashed lines 
indicate the simulated analogs to the Magellanic Clouds, defined 
here as subhalos with $\vmax > 60~\kms$, which we remove from our 
analysis and plot only for illustrative purposes.  The dotted lines 
indicate circular velocity profiles that fall below the $1\sigma$ 
error on $\vhalf$ for at least one of the MW dSphs~--~these subhalos 
are nominally consistent with the observational data and can host 
a MW satellite.  The solid lines, however, have circular velocities
that lie above \textit{all} the dSphs and therefore qualify as 
``massive failures"~--~subhalos without observational counterparts.  
Nearly all of these massive failures are large enough, even today,
to have formed stars in the presence of an ionizing background 
\citep{Bullock2000,Somerville2002,Sawala2014}.

Though the TBTF problem remains evident in all three models plotted 
here,
    \footnote{Though we do not plot it, the central halo in the \litep\ 
	cosmology hosts eight massive failures.} 
the number of massive failures is noticeably reduced in the \bicep\ 
cosmology relative to the \planck\ model.  Perhaps surprisingly, the 
running power spectrum of \bicep\ eliminates more massive failures 
than the chosen WDM free-streaming cutoff.\footnote{Though a lighter 
	WDM mass will be more effective \citep[e.g., ][]{Schneider2014}, 
	it is constrained by the Ly-$\alpha$ forest \citep{Viel2013} and 
	subhalo counting \citep{Polisensky2011}; as discussed in 
	Section~\ref{sec:intro}, however, these constraints are subject 
	to systematic uncertainties that are currently difficult to quantify.}
Moreover, the remaining massive failures in the \bicep\ model lie well below the equivalent 
curves in the \planck\ cosmology, which acts to increase the efficacy of other 
processes (e.g. supernovae feedback) that may further reduce the central densities.  
Similarly, the \bicep\ cosmology significantly lowers the number of subhalos 
that are consistent with only Draco and Ursa Minor, the two highest density 
galaxies in the sample.  Overall, the \bicep\ cosmology significantly reduces 
the magnitude of the TBTF problem, even without invoking baryonic processes
that may further reduce the central densities \citep[e.g.][]{Zolotov2012},
perhaps in a cosmology-dependent manner.

In addition to reducing the number of massive failures, the increase in $\rmax$
at fixed $\vmax$ in the \bicep\ cosmology implies a reduction in the substructure
boost, i.e., the expected dark matter annihilation signal from subhalos.  As noted
above, the signal from a single halo scales as 
$\rho_{\rm s}^2r_{\rm s}^3 \propto \vmax^4/\rmax$.  Therefore, an increase
of $35\%$ in $\rmax$ at fixed $\vmax$ directly results in a $35\%$ reduction 
in the annihilation signal.  Furthermore, the overall boost is obtained by summing 
the signal from all the substructure by integrating the mass (or $\vmax$) function to 
masses well below $\msun$ \citep{Martinez2009}; assuming that the $\sim50\%$ 
offset in the $\vmax$ function at the resolution limit ($\vmax = 8~\kms$) remains 
constant at lower masses, this implies that the substructure boost in the \bicep\ 
cosmology may be a factor of $\sim5$ lower than in \planck.  Moreover, the 
increasing roll-off of $P(k)$ at small scales implies that the relative offsets in both the 
$\vmax$ function and the $\rmax-\vmax$ relationship are even larger at small 
masses; the estimate will realistically be larger than $5$.

\begin{figure}
\includegraphics[width=\columnwidth]{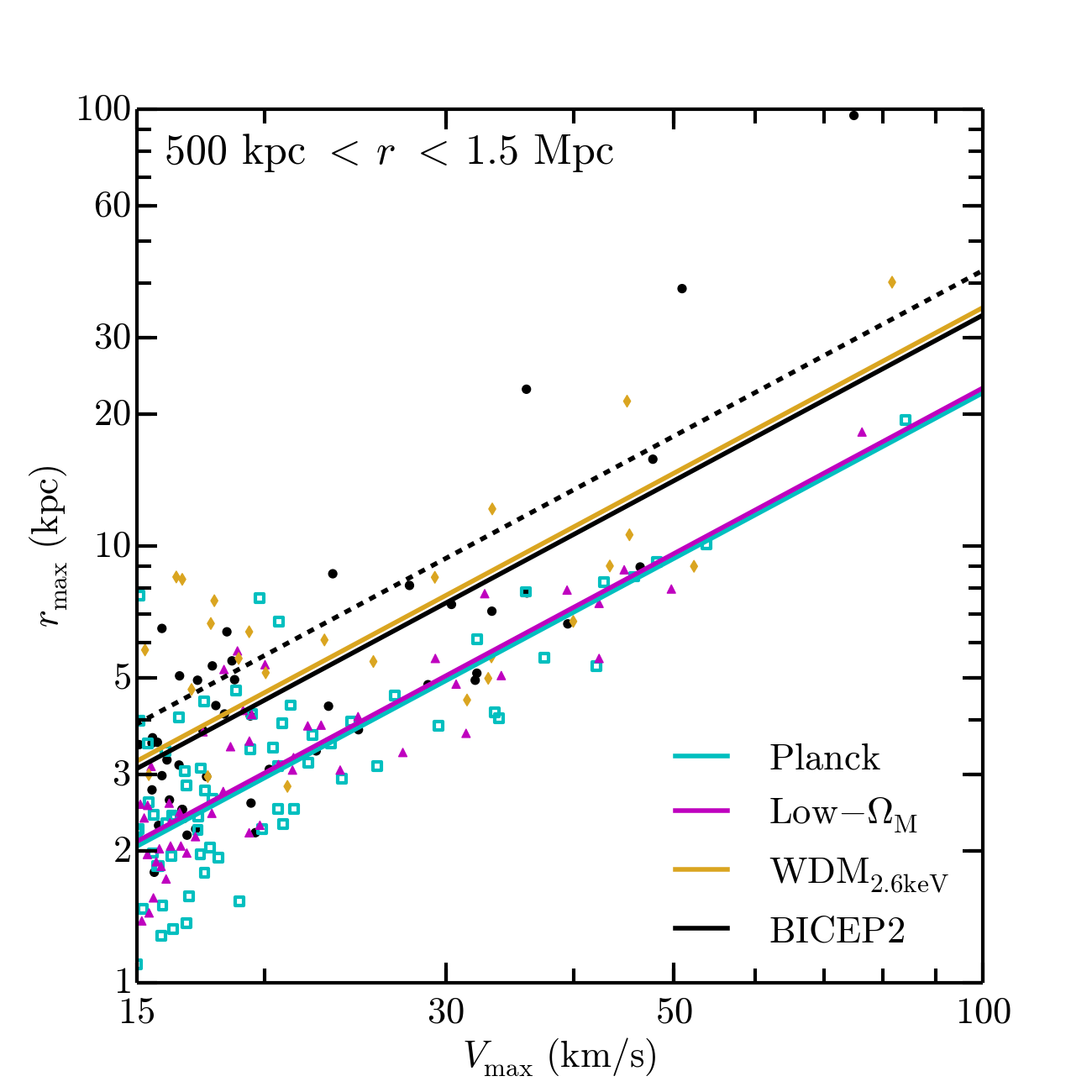}
\caption{The $\rmax$-$\vmax$ relation for halos in the fields
around the zoom-in target.  Solid lines again plot fits 
(Equation~\ref{eqn:rmaxfit}) weighted by $\vmax$, with the 
log-slope again held fixed at the best-fit value for \planck,
$p = 1.26$.  The normalizations, $A$, are 1.23 (\planck), 1.26 (\litep),
1.93 (\wdmp), and 1.85 (\bicep)~--~suppression in $P(k)$ at small 
scales in \wdmp\ and \bicep\ results in normalizations $\sim50\%$ lower.
The most massive halo in the \bicep\ field is excluded from the fit because 
the anomalously high $\rmax$ is due to an ongoing merger~--~including 
that halo results in a $20\%$ larger normalization ($A = 2.35$), which is 
plotted as a black dashed line.}
\label{fig:fieldrmaxvmax}
\end{figure}

For subhalos, the $\rmax-\vmax$ relation is due to a combination of the 
concentration-mass relationship at the time of formation and tidal 
stripping after infall onto the central host \citep{Bullock2001,Ludlow2014}.  
To more directly probe the former, Figure~\ref{fig:fieldrmaxvmax} plots 
$\rmax$ and $\vmax$ for halos in the field surrounding the central host, 
along with power-law fits (Equation~\ref{eqn:rmaxfit}) with $p$ again
held fixed at best fit value in the \planck\ simulation, $p = 1.26$.  We limit 
ourselves to objects at least $500~\kpc$ from the central host to avoid 
the majority of ``backsplash" galaxies that have interacted with the 
host in the past \citep{Teyssier2012,ELVIS}, which may have undergone 
significant tidal stripping, and  we select halos within  $1.5~\mpc$ to 
avoid high mass (low resolution) contaminating particles.  

While the agreement between the \planck\ and \litep\ models remains
in the field (as expected due to their similar power spectra), the 
effects of the modifications to $P(k)$ are apparent in both the \wdmp\ 
and \bicep\ simulations.  The latter two display significantly lower
density halos, consistent with the suppression in  power spectra at the
time of formation; the fits to both are $\sim50\%$ higher than the fit
in the \planck\ cosmology.  The most massive nearby field halo in the
\bicep\ simulation is undergoing a major merger, resulting in an 
anomalously large $\rmax$ and we therefore perform the fit with
and without that object.  Including it results in the fit plotted as a black
dashed line; the fit without that point is plotted as a solid black line.

\section{Conclusions}
\label{sec:conclusions}
We have tested the impact of the suppressed small-scale primordial power spectrum 
suggested by the recent \bicep\ results by simulating structure formation
both with the ``running" power spectrum suggested by these results and 
with the cosmology suggested by the \planck\ experiment, and using two control 
models~--~the \planck\ model with a free-streaming cut-off corresponding to a 
WDM particle mass of 2.6~keV (thermal) and the \planck\ power spectrum with an 
artificially lowered $\om$.  We have simulated the evolution of identical 
$(50~\hmpc)^3$ volumes from $z = 125$ until $z = 3$ and the formation of a 
MW-size host until $z = 0$ at high resolution.  These simulations indicate 
that the suppression in the primordial power spectrum at small scales results 
in mild offsets in the large-scale halo mass function (consistent with 
expectations from linear theory) and non-trivial differences in the subhalo 
$\vmax$ function and the inner structure of both field and satellite halos.  
Specifically:

\begin{itemize}
\item The  $\vmax$ function of subhalos around a MW-size host in
	the \bicep\ cosmology lies well below that of the same host in the \planck\
	model for $\vmax \lesssim 40~\kms$, even after normalizing for the 
	differing sizes of the hosts.  There are twice as many resolved 
	($\vmax > 8~\kms$) subhalos within the virial radius of the central host 
	in the \planck\ simulation as result in the \bicep\ cosmology.  The \planck\
	and \litep\ models agree after scaling for the host mass.  Unsurprisingly,
	the \wdmp\ simulation results in only $\sim10\%$ as much substructure as 
	our fiducial \planck\ run.
	
\item Although masses of the largest subhalos around our selected host appear 
	to be mostly unaffected by the changes in cosmology, the average concentrations
	(quantified here by the relationship between $\rmax$ and $\vmax$) of subhalos 
	are significantly lower in the \bicep\ cosmology than any of the \planck-like
	models and our \wdmp\ run.  This increase in $\rmax$ at fixed $\vmax$ 
	alleviates the too-big-to-fail problem, and may increase the 
	efficacy of baryonic processes that could further reduce the central densities.  
		
\item Taken together, the above two results imply that the substructure	``boost," 
	the contribution to the dark matter annihilation signal due to subhalos, is at
	least a factor of $\sim5$ times smaller in the \bicep\ cosmology.  Although the 
	absolute value of the boost depends on many assumptions and is an uncertain 
	quantity, this relative modification should be more robust and will work to lower 
	previous upper limits to order unity.
\end{itemize}

While the above conclusions are drawn from simulations of only a single
MW-size host halo, the overall trends demonstrated should hold for all
such systems.  Though there is significant scatter between MW-size systems
\citep[e.g.][]{Boylan-Kolchin2010}, the relative offset from the mean in 
the substructure population of a single host appears to remain largely static
across cosmologies \citep{Horiuchi2014}.  Therefore, the precise magnitude
of the above changes may vary, but the general result that subhalos are less
numerous and less dense in the \bicep\ model compared to \planck\ is robust.
In order to accurately determine the range of substructure suppression and
changes in concentration, one requires a large sample of simulations similar 
to those shown here; we elect to instead illustrate the general trends only.

Our results indicate that the level of spectral index running that reconciles the \bicep\ 
measurement with other constraints has interesting effects on dark matter structure 
over a range of scales.  These changes are most evident at the smallest scales, where they 
help to alleviate small-scale issues with CDM.  Though not addressed here, this type of 
reduction in small-scale power could have interesting implications for understanding 
cosmic reionization, which may require the early collapse of small halos and thus a fair 
amount of power on $\sim 10^8 \msun$ scales \citep[e.g.][]{somerville2003,Robertson2013}, 
and conversely studies of the early Universe may constrain the allowed 
running \citep[similar to the constraints placed on WDM by][]{schultz2014}.  Signs of a 
non-trivial primordial power spectrum may also be explored in the Ly-$\alpha$ forest. 

While it should be noted that inflationary models with precisely constant running at the 
level we have investigated have difficulty producing enough $e$-foldings \citep{Easther2006}
and likely have higher order corrections to the power spectrum in this parameterization 
\citep{Abazajian:2005dt}, there are feasible models with scale-dependent running that produce 
similar suppression of power at dwarf scales to that considered here \citep[e.g.,][]{Kobayashi2011,Wan2014}.
The broad point of this work is to highlight the salient role that a non-trivial primordial 
power spectrum has in affecting small-scale predictions in \lcdm. In light of the exciting 
\bicep\ results interpreted as evidence for inflationary gravitational waves, the need to 
consider non-standard primordial power spectra in structure formation studies has grown 
all the more urgent.

\vskip1cm

\noindent {\bf{Acknowledgments}} \\
The authors thank Quinn Minor, Mike Boylan-Kolchin, Amjad Ashoorioon,
Daniel Figueroa, and particularly the anonymous referee for helpful
comments.  SGK and JSB were partially supported by NSF grants 
AST-1009973 and AST-1009999.  SH is supported by the JSPS fellowship 
for research abroad, KA is supported by NSF CAREER Grant No.\ PHY-11-59224,
and MK is supported by NSF Grant Nos. PHY-1214648 and PHY-1316792.

We also acknowledge the computational support of the NASA Advanced 
Supercomputing Division and the NASA Center for Climate Simulation, 
upon whose \textit{Pleiades} and \textit{Discover} systems the 
simulations were run, and the \textit{Greenplanet} cluster at UCI, 
upon which much of the secondary analysis was performed.
This research has made use of NASA's Astrophysics Data System.

\bibliographystyle{mn2e}
\bibliography{bicep}

\label{lastpage}
\end{document}